\documentstyle[12pt,fleqn]{article}
\begin{document}
\newcommand{\kslash}{\mbox{$\displaystyle\not\mkern-4mu k$}}
\newcommand{\Dslash}{\mbox{$\displaystyle\not\mkern-4mu D$}}
\newcommand{\be}{\begin{equation}}
\newcommand{\ee}{\end{equation}}
\newcommand{\ds}{\displaystyle}
\mbox{ }\hfill{\normalsize ITP-92-28E}\\
\mbox{ }\hfill{\normalsize March 1992}
\begin{center}
{\Large \bf Quantum Solitons with Cylindrical Symmetry} \\[.5cm]
\vspace{1cm}
{\large N.~Chepilko\footnote{Institute of Physics,
Academy of Sciences of Ukraine, Kiev, Ukraine}, A.~Kobushkin,
A.~Syamtomov }\\[.5cm]
{\large \it N.N.Bogolyubov Institute for Theoretical Physics, \\
Ukrainian Academy of Sciences, Kiev 143, Ukraine
}
\end{center}
\date{}
\vspace{.5cm}
\begin{abstract}
Soliton solutions with cylindrical symmetry are investigated within the
nonlinear $\sigma $-model disregarding the Skyrme-stabilization term.
The solitons are stabilized by quantization of collective breathing
mode and collapse in the $\hbar \rightarrow 0$ limit. It is shown that for
such stabilization mechanism the model, apart from  solitons  with  integer
topological number $B$, admits the solitons with half-odd $B$. The  solitons
with integer $B$ have  standard  spin-isospin  classification,
while $B={\ds {1\over 2}}$ solitons are shown to be characterized by spin,
isospin and some additional "momentum".
\end{abstract}
\newpage
\section{Introduction}
\indent
Few years ago a new class of  spherically-symmetrical  solitons
in the nonlinear $\sigma$-model, whose stability is provided on the  basis
the quantization of  collective  coordinates  chosen  suitably  was
shown in [1-3]. Such solitons does not exist when the Planck  constant
tends to zero and in this sense they are called quantum  solitons.\\
\indent
On the other hand, in the framework of the standard Skyrme model [4]
(the model, where solitons are stabilized by a fourth-order
derivative term added by hand) a  spherically-symmetrical  hedgehog
configuration was shown to have the lowest energy only in a  sector
with topological number ${\bf B}=1.$ For example, for the systems with  the
baryon number 2 an Ansatz with the  cylindrical  symmetry  is  more
preferable and was studied by many authors [5-7].  The  purpose  of
this paper is to explore the  cylindrically-symmetrical  Ansatz  in
the theory of quantum solitons.\\
\indent
We are starting from the action of the standard $SU(2)$ nonlinear
$\sigma$-model:
\be
W=-{f^{2}_{\pi}\over 4} \int d^{4}xTr(L_{\mu }L^{\mu }),\ \
\ L_{\mu}=U^{+}\partial _{\mu }U,
\label{eq1.1}
\ee
where $x^{\mu}=(t,\vec{x})$ is the Minkowski space coordinate, $U=U(t;\vec{x})$
is the $(2\times 2)$ chiral field matrix,  $f_{\pi }$  is  the  pion  decay
constant  and $L^{\mu}=\eta^{\mu \nu}L_{\nu}$ with the metric tensor
$\eta ^{\mu \nu}$ of Minkowski space;  the  space signature is taken
to be $(+ - - -)$.\\
\indent
According to the Derrick theorem [8] the static soliton  configuration
does not exist in the model with the action (\ref{eq1.1}). But for
a time-dependent configuration the conditions of the Derrick  theorem are
not fulfiled and therefore  we  start  from  the  following
Ansatz, which is explicitly time-dependent:
\be
U(t;r,\varphi ,\vartheta )=e^{-i{m\over 2}\varphi _{3}\tau_{3}}\exp
\left\{iF_{a}\left({r\over {\lambda (t)}},\vartheta \right)
\tau _{a}\right\}e^{i{m\over 2}\varphi _{3}\tau _{3}},
\label{eq1.2}
\ee
where $r, \varphi , \vartheta $ are the usual spherical coordinates,
$\tau_{a} (a=1,2,3)$ are the isospin Pauli matrices, $m$ is some integer
number and $\lambda (t)$ is  a configuration size parameter; the latter
will be considered in this  paper as a collective ("breathing") coordinate.\\
\indent
In Ref.7 the static limit $(\lambda (t)=1)$ of the Ansatz (\ref{eq1.2})
was applied
for the configurations with $m=2$ and ${\bf B}=2$ in the  framework  of
the Skyrme model. From the Ansatz (\ref{eq1.2}) one can also simply  obtain
the hedgehog configuration by using here the static limit and  setting
$m=1$ and $F_{1}=F(r)\sin \vartheta ,\ F_{2}=0,\ F_{3}=F(r)\cos \vartheta $.\\
\indent
The outline of the paper is as follows. Taking into account the
noncommutativity of the breathing coordinate $\lambda (t)$ and the  velocity
$\dot{\lambda }=\frac{\textstyle d\lambda (t)}{\textstyle dt}$ we derive an
effective Lagrangian for breathing   motion  in Sec.II. It is shown that
due to the uncertancy principle  a  repulsive term is generated, what
prevents a collapse of the chiral field. To define the chiral field
functions $F_{a}$ for arbitrary $m$ a  system of coupled differential
equations  in  partial derivatives  is obtained. We demonstrate that for
the cylindrically-symmetrical configurations there appear solitons with
integer as well as half-odd topological charge. In Sec.III the quantization
of spatial and isospatial rotations of soliton is considered. For solitons
with integer topological charge the usual quantum numbers (spin and
isospin) are obtained, meanwhile for solitons  with  ${\ds 1\over
\ds 2}$-topological  charge there appears additional conserved "momentum"
operator, which  commutes with spin and isospin operators (Sec.IV). This
momentum $\zeta $  is shown to be integer and is  restricted by $|j-t|\le
\zeta \le j+t$, where $j$ and $t$ are spin and isospin, respectively.
Conclusions and some speculations related to status of the quantum
solitons  with  cylindrical symmetry are given in the last section.\\
\section{Basic Equations}
\subsection{Effective Lagrangian}
\indent
Following the approach of Ref.1, which  was  developed  in  the
theory of spheri\-cally-symmetrical solitons, we are postulating that
the generalized coordinate $\lambda $ and velocity $\dot{\lambda }$ do not
commute:
\be
[\lambda ,\dot{\lambda }]=i\hbox{{\it f}}(\lambda ),
\label{eq2.1}
\ee
where $f(\lambda )$ is to be determined  after  the  canonical
quantization condition is required.\\
\indent
To obtain an effective potential for the breathing motion it is
useful to replace the Minkowski coordinates $(x^{\mu })$ by a new one
$(z^{\alpha })$ according to
\be
(x^{\mu })\equiv (t,\vec{x} )=(\tau,\lambda (\tau)\vec{z}),\ \
U(t,\vec{x})=\hat{U}(\vec{z}).
\label{eq2.2}
\ee
Thus one goes from the representation, where the  configuration
vibrates in Min\-kow\-ski space, to a new one, where the  configuration
is "static", but quantum vibrations are transferred to the  geometrical
structure of the adopted space.\\
\indent
After simple calculations (see, e.g., Ref.1) and  taking  into
account (\ref{eq2.1}) one obtains the following  effective  Lagrangian  for
the breathing motion:
\begin{eqnarray}
W & = & \int \Lambda \, d\tau, \nonumber \\
\Lambda & = & \sigma _{2}[F,\Theta
]\left(\dot{\lambda } \lambda \dot{\lambda }-ff^{\prime }-{5\over 4}\lambda
^{-1} f^{2}\right)-\sigma _{1}[F,\Theta ]\lambda ,
\label{eq2.3}
\end{eqnarray}
where $f^{\prime }=\frac{\textstyle df(\lambda )}{\textstyle d\lambda }$
and $\sigma _{1}[F,\Theta ]$ and $\sigma _{2}[F,\Theta ]$ are the following
functionals:
\begin{eqnarray}
\sigma _{1}[F,\Theta ] & = & \frac{f^{2}_{\pi }}{2}\int d^{3}z
\left[(\vec{\nabla}F)^2+\sin ^{2}F(\vec{\nabla}n_a)(\vec{\nabla}n_a)\right],
\label{eq2.4a} \\
\sigma _{2}[F,\Theta ] & = & \frac{f^{2}_{\pi }}{2}\int d^{3}z\,
z_{i}z_{j}\left[(\nabla _{i}F)(\nabla _{j}F)+\sin ^{2}F(\nabla _{i}n_{a})
(\nabla _{j}n_{a})\right].
\label{eq2.4b}
\end{eqnarray}
Here functions $F=F(z,\vartheta ),\ n_{a}=n_{a}(z,\vartheta ),\ a=1,2,3$
parametrize the chiral field $\hat{U}(\vec{z})$ by
\begin{eqnarray}
F_{1} & = & F\cos \Theta,\ \ F_{2}=0,\ \ F_{3}=F\sin \Theta,
\label{eq2.5a} \\
n_{1} & = & \cos m\varphi  \cos \Theta ,\ \ n_{2}=\sin m\varphi  \cos \Theta,
\ \ n_{3}=\sin \Theta ,
\label{eq2.5b}
\end{eqnarray}
where $\Theta =\Theta (z,\vartheta ),\ F=F(z,\vartheta ),\ z=|\vec{z}|$
and $\nabla _{i}={{\ds \partial } \over {\ds {\partial z^i}}}$.\\
\indent
When using the transformation $(\lambda ,\dot{\lambda })
\rightarrow (\xi ,\dot{\xi })$, where
\be
\lambda =\sigma _{1}[F,\Theta ]^{-1}\xi ^{2/3} ,\ \ \dot{\lambda}=
\left\{\dot{\xi }, {\partial \lambda\over \partial \xi }
\right\},
\label{eq2.6}
\ee
and $\{a,b\}={{\ds 1}\over {\ds 2}}(ab+ba)$, the
Lagrangian (2.3) is reduced to
\begin{eqnarray}
\Lambda ={{\ds 1}\over {\ds 2}}M[F,\Theta ]\dot{\xi}^{2}-v(\xi ),
\ \ v(\xi )=\xi ^{2/3}+\frac{1}{8M[F,\Theta ]}\xi ^{-2},
\label{eq2.7} \\
M[F,\Theta ] ={\ds 8\over \ds 9}{{\ds {\sigma _{2}
[F,\Theta ]}}\over {\ds {\sigma ^{3}_{1}[F,\Theta ]}}}. \nonumber
\end{eqnarray}
To obtain (\ref{eq2.7}) one
has to put $f(\lambda )=(2\sigma _{2}\lambda )^{-1}$, which is
consistent with the canonical quantization condition $[\xi ,P_{\xi }]=i$,
where $P_{\xi }$  is  a momentum, conjugative to $\xi $ .\\
\indent
It must be emphasized here that the Lagrangian  (\ref{eq2.7})  is  invariant
under the scale transformation
\be
F(z,\vartheta )\rightarrow F_{\beta }(z,\vartheta )\equiv F(\beta z,\vartheta
),
\ \ \Theta (z,\vartheta )\rightarrow \Theta _{\beta }(z,\vartheta )\equiv
\Theta (\beta z,\vartheta ) .
\label{eq2.8}
\ee
The role of this invariance of an effective Lagrangian in theory of
quantum solitons was discussed in Ref.3.\\
\subsection{Variational Principle and Differential Equations}
\indent
The chiral field for the cylindrically-symmetric  configuration
is defined by the same variational equation, which  was derived for
spherically-symmetric chiral field
\be
3\frac {\delta \sigma _{1}}{\sigma _{1}}-\frac{\delta \sigma _{2}}
{\sigma _{2}}=0,
\label{eq2.9}
\ee
where functionals $\sigma _{1}$ and $\sigma _{2}$ are defined by
(\ref{eq2.4a}, \ref{eq2.4b}).\\
\indent
According to a prescription of Refs.1,3 we  shall  introduce  a
new variable $\vec{y}$ and functions $\tilde{F}(\vec{y})$ and
$\tilde{\Theta }(\vec{y})$  instead  of  the  old ones:
\begin{eqnarray}
\vec{y} & = & z^{-1}_{0}\vec{z},\ \ z^{2}_{0}=3\frac{\sigma _{2}[F,\Theta ]}
{\sigma _{1}[F,\Theta ]},
\label{eq2.10a} \\
\tilde{F}(\vec{y}) & = & F(z_{0}\vec{y}),\ \ \tilde{\Theta }(\vec{y})
=\Theta (z_{0}\vec{y}).
\label{eq2.10b}
\end{eqnarray}
Calculating in (\ref{eq2.9}) the variations over $\delta F$ one obtains  the
first
field equation
\be
-\partial _{i}\partial ^{i}\tilde{F}+\partial _{i}\left(y^{i}y^{j}\partial _{j}
\tilde{F}\right)+\sin \tilde{F}\cos \tilde{F}\left(\partial _{i}\tilde{n}_{a}
\partial _{i}\tilde{n}_{a}-y^{i}y^{j}\partial _{i}\tilde{n}_{a}
\partial _{j}\tilde{n}_{a}\right)=0,
\label{eq2.11a}
\ee
where $\partial _{k}=\frac{\textstyle \partial }{\textstyle \partial
y_{k}}$ and $\tilde{n}_{a}(\vec{y})=n_{a}(z_{0}\vec{y})$.\\
\indent
The second equation can be  derived  from  Eq.(\ref{eq2.9})  using  the
standard variation procedure over $\delta \tilde{n}_{a}$ with the
constraint $\tilde{n}_{a}\tilde{n}_{a}=1$:
\begin{eqnarray}
\lefteqn{2\cot \tilde{F}\partial _{i}\tilde{F}\left(\partial _{i}
\tilde{n}_{a}-
y^{i}y^{j}\partial _{j}\tilde{n}_{a}\right) + \partial _{j}
\left(\partial _{i}\tilde{n}_{a}-y^{i}y^{k}\partial _{k}\tilde{n}_{a}\right)}
\nonumber \\
& & +\left(\partial _{i}\tilde{n}_{b}\partial _{i}\tilde{n}_{b}-
y^{i}y^{k}\partial _{i}\tilde{n}_{b}\partial _{k}\tilde{n}_{b}\right)
\tilde{n}_{a}=0
\label{eq2.11b}
\end{eqnarray}
with $\tilde{n}_{a}\tilde{n}_{a}=1$ and $\tilde{n}_{a}\partial _{i}
\tilde{n}_{a}=0$.\\
\indent
Apparently, for the hedgehog-like  configuration  of
Refs.1-3 Eq.(\ref{eq2.11b}) is satisfied identically, while
Eq.(\ref{eq2.11a}) is reduced to the field equation of Ref.1.\\
\indent
One can  rewrite Eqs.(\ref{eq2.11a}, \ref{eq2.11b}) in spherical  coordinates
$y=|\vec{y}|,\ \vartheta $ and $\varphi $:
\begin{eqnarray}
\lefteqn{{\partial \over \partial y}\left[\left(y^{4}-y^{2}\right)
{\partial \tilde{F}\over \partial y}\right]-{1\over \sin \vartheta }
{\partial \over \partial \vartheta }\left[\sin \vartheta
{\partial \tilde{F}\over \partial \vartheta }\right] +} \nonumber \\
& & \sin \tilde{F} \cos \tilde{F}\left[\left(y^{2}-y^{4}\right)\left[
{\partial \tilde{\Theta }\over \partial y}\right]^{2}\right. + \left.
\left[{\partial \tilde{\Theta }\over \partial \vartheta }\right]^{2}+
m^{2}\left[{\sin \tilde{\Theta }\over \sin \vartheta }\right]^{2}\right]=0,
\label{eq2.12a} \\
\lefteqn{\sin \vartheta {\partial \over \partial y}\left[\left(y^{4}-
y^{2}\right) \sin ^{2}\tilde{F}{\partial \tilde{\Theta }\over \partial y}
\right] - {\partial \over \partial y}\left[\sin \vartheta \sin ^{2}
\tilde{F}{\partial \tilde{\Theta }\over \partial y}+\right. } \nonumber \\
& & + \left. {m^{2}\over 2}\sin ^{2}\tilde{F}
{\sin 2\tilde{\Theta }\over \sin \vartheta }\right]=0.
\label{eq2.12b}
\end{eqnarray}
Both equations have three singular points: $y=0,\ y=1$ and $y=\infty $ and
for further analysis it is important to know the  behavior  of  the
functions $\tilde{F}(y,\vartheta )$ and $\tilde{\Theta }(y,\vartheta )$ near
these points.\\
\subsection{Behavior at Singular Points}
\indent
Suggesting  the  analytical  properties  of  the  solutions   of
Eqs.(\ref{eq2.12a},\ref{eq2.12b}) one has to expand the functions
$\tilde{F}(y,\vartheta )$ and $\tilde{\Theta }(y,\vartheta )$  near the origin
$y=0$ as
\begin{eqnarray}
\tilde{F}(y,\vartheta ) & = & F_{0}+y^{\mu
}F_{1}(\vartheta )+\ldots ,\ \ \ \mu >0, \label{eq2.13a} \\
\tilde{\Theta
}(y,\vartheta ) & = & \Theta _{0}(\vartheta )+y^{\nu }\Theta _{1}(\vartheta )+
\ldots,\ \ \ \nu >0,
\label{eq2.13b}
\end{eqnarray}
where $\mu $ and $\nu$ have  to  be  determined  from  the  field  equations
(\ref{eq2.12a},\ref{eq2.12b}). Putting (\ref{eq2.13a},\ref{eq2.13b}) into
Eq.(\ref{eq2.12a}) one easily obtains at  zero order of $y$-expansion:
\be
\sin F_{0}\cos F_{0}\left[\left({\partial \Theta _{0}\over \partial \vartheta }
\right)^{2}+m^{2}\left({\sin \Theta _{0}\over \sin \vartheta }\right)^{2}
\right]=0.
\label{eq2.14}
\ee
To satisfy this equation we should take
\be
F_{0}={\pi \over 2}n,\ \ n=0,\pm 1,\pm 2,\ldots.
\label{eq2.15}
\ee
The topological charge for the solitons is defined as usual
\be
{\bf B}=\int d^{3}x \, B^{0}(x),\ \ {where}\ \ B^{\mu }(x)={1\over 24\pi ^{2}}
\epsilon ^{\mu \nu \rho \sigma }Tr(L_{\nu }L_{\rho }L_{\sigma }), \nonumber
\ee
while $L_{\nu }$ was defined in (\ref{eq1.1}). Simple calculations give ${\bf
B}={{\ds nm}\over {\ds 2}}$  and the field equations
(\ref{eq2.12a},\ref{eq2.12b}) admit solitons with integer and half-odd
topological charge. The latter  does not  exist in  the  standard Skyrme model
[4].  Introducing a variable $\chi \equiv m\ln \left(\tan {{\ds \vartheta}
\over {\ds 2}}\right)$ one  obtains the sine-Gordon equation for $\Theta _{0}
(\vartheta (\chi ))$  at the zero order  of $y$-expansion of
Eq.(\ref{eq2.12b}):
\be
{d^{2}\Theta _{0}\over d\chi ^{2}}={\sin 2\Theta _{0}\over 2},
\label{eq2.16}
\ee
which has the well known solution
\be
\Theta _{0}=2\tan ^{-1}\left(e^{\chi }\right)=2\tan ^{-1}\left(\tan
{\vartheta \over 2}\right)^{m}.
\label{eq2.17}
\ee
The next order of the $y$-expansion of Eq.(\ref{eq2.12a}) gives  an
equation for $F_{1}(\vartheta )$:
\be
{d\over dz}\left[(1-z^{2}){dF_{1}\over dz}\right]+\left[\mu (\mu +1)-(-1)^{n}
8m^{2}{(1+z^{2})^{m-1}\over [(1+z)^{m}+(1-z)^{m}]^{2}}\right]F_{1}=0,
\label{eq2.18}
\ee
where $z$ denotes $\cos \vartheta $, and $n$ is determined by (\ref{eq2.15}).
\\ \indent
At zero order of $(y-1)$-expansion of Eqs.(\ref{eq2.12a},\ref{eq2.12b})
one gets
\begin{eqnarray}
\lefteqn{{\partial F(1,\vartheta )\over \partial y} ={1\over 2\sin \vartheta }
{d\over d\vartheta }\left[\sin \vartheta {dF(1,\vartheta )\over
d\vartheta }\right]} \nonumber \\
& &-{\sin 2F(1,\vartheta )\over 4}\left[\left(
{d\Theta (1,\vartheta )\over d\vartheta }\right)^{2}\right. +\left.
m^{2}\left({\sin \Theta (1,\vartheta )\over \sin \vartheta }\right)^{2}\right],
\label{eq2.19a} \\
\lefteqn{{\partial \Theta (1,\vartheta )\over \partial y} =\frac{{\ds d\over
\ds d\vartheta }\left[\sin ^{2}F(1,\vartheta )\sin \vartheta
{\ds d\Theta (1,\vartheta )\over \ds d\vartheta }\right]}{2\sin \vartheta
\sin ^{2}F(1,\vartheta )}-m^{2}\left[{\sin 2\Theta (1,\vartheta )\over
4\sin ^{2}\vartheta }\right].}
\label{eq2.19b}
\end{eqnarray}
Finally, at $y\rightarrow \infty $ the functions $F$ and $\Theta $ behave as
\begin{eqnarray}
F(y,\vartheta ) & = & y^{-3}f_{0}(\vartheta )+y^{-5}f_{1}(\vartheta )+\ldots,
\label{eq2.20a} \\
\Theta (y,\vartheta ) & = & \theta _{0}(\vartheta )+y^{-2}\theta _{1}
(\vartheta )+\ldots,
\label{eq2.20b}
\end{eqnarray}
where the relationships between functions in this  expressions  are
given by
\begin{eqnarray}
f_{1}(\vartheta ) & = & {3\over 5}f_{0}+\frac{{{\ds d}\over
\ds {d\vartheta }}\left[\sin \vartheta {\ds df_{0}\over \ds d\vartheta }
\right]}{10\sin \vartheta }-f_{0}\left[\left({d\Theta _{0}\over d\vartheta }
\right)^{2}+m^{2}\left({\sin \Theta _{0}\over \sin \vartheta}\right)^{2}
\right],
\label{eq2.21a} \\
\theta _{1}(\vartheta ) & = & \frac{{\ds d\over \ds d\vartheta }\left[
\sin \vartheta {\ds d\Theta _{0}\over \ds d\vartheta }\right]}{10f^{2}\sin
\vartheta }-m^{2}{\sin 2\Theta _{0}\over 20\sin ^{2}\vartheta }.
\label{eq2.21b}
\end{eqnarray}
\\
\section{Spatial and Isospatial Rotations}
\indent
 From the configuration considered in the previous  section  the
infinite set of configurations, all degenerate in  energy,  can  be
constructed by spatial and isospatial rotations:
\be
\hat{U}(\vec{z},A,B)=AU\left[D(B)\vec{z}\right]A^{+},
\label{eq3.1}
\ee
where $D^{i}_{j}(B)={\textstyle 1\over 2}Tr\left[\tau_{i}B\tau_{j}B^{+}
\right]$ is an orthogonal matrix of spatial rotation given by a $SU(2)$
matrix $B$; an isospatial  rotation  of  the chiral field $U$ is given
by a $SU(2)$ matrix $A$.  The usual way to quantize the rotating degrees
of freedom is to promote the  matrices  $A$ and $B$ to collective
coordinates $A(t)$ and $B(t)$.\\
\indent
Here one comment must be given. Due to the fact that the theory
in question gives no stable soliton solution in the classical limit
$\hbar \rightarrow 0$ the traditional semi-classical quantization scheme
became  nonapplicable here. But an example of a vibrating and rotating quantum
soliton for the hedgehog-like configuration $U(\vec{z})$ [3] shows that the
contribution of a rotational energy of the field is less  than  one
quarter of total soliton energy. Therefore we shall assume  further
that the chiral field matter distribution is that for the  nonrotating
soliton considered in Sec.II and the total soliton  energy  is given by
\be
E=E_{0}+T,
\label{eq3.2}
\ee
where $E_{0}$ is an energy of  the  nonrotating  soliton,  which  is  an
eigenvalue of the Hamiltonian corresponding to the  effective  Lagrangian
$\Lambda $ for the vibrating motion (Eq.(\ref{eq2.3})):
\be
\hat{H}\Psi =E_{0}\Psi,\ \ \hat{H}=-{1\over 2M[F,\Theta ]}{d^{2}
\over d\xi ^{2}} +v(\xi ) ;
\label{eq3.3}
\ee
$T$ is the kinetic energy related to rotations. It is  determined  by
the Hamiltonian [7]:
\be
\hat{H}_{rot}={1\over 2}a_{i}U_{ij}a_{j}-a_{i}W_{ij}b_{j}+
{1\over 2}b_{i}V_{ij}b_{j},
\label{eq3.4}
\ee
where $a_{j}=-iTr[\tau_{j}A^{+}\dot{A}],\ b_{j}=iTr[\tau_{j}\dot{B}B^{+}],
\ \dot{A}\equiv  {{\ds dA(t)}\over {\ds dt}},\ \dot{B}\equiv
{\ds dB(t)\over \ds dt}$  and  the inertia tensors are expressed in
terms of chiral field $U\equiv U(\vec{z})$ by
\begin{eqnarray}
U_{ij} & = & {\xi ^{2}\over
\sigma ^{3}_{1}[F,\Theta ]}u_{ij},
\label{eq3.5a} \\
W_{ij} & = & {\xi ^{2}\over \sigma ^{3}_{1}[F,\Theta ]}w_{ij},
\label{eq3.5b} \\
V_{ij} & = & {\xi ^{2}\over \sigma ^{3}_{1}[F,\Theta ]}v_{ij},
\label{eq3.5c}
\end{eqnarray}
where
\begin{eqnarray}
u_{ij} & = & {1\over 4f_{\pi }}\int
d^{3}z Tr\left(U^{+}\left[{1\over 2}\tau_{i}, U\right]U^{+}\left[{1\over
2}\tau_{j},U\right]\right),
\label{eq3.6a} \\
w_{ij} & = & -u_{ij}\left\{\left[{1\over 2}\tau_{j},U\right] \rightarrow
-i(\vec{r}\times \vec{\nabla})_{j}U\right\},
\label{eq3.6b} \\
v_{ij} & = & -w_{ij}\left\{\left[{1\over 2}\tau_{i},U\right] \rightarrow
-i(\vec{r}\times \vec{\nabla})_{i}U\right\}.
\label{eq3.6c}
\end{eqnarray}
Note that the inertia tensors $U_{ij},\ W_{ij},\ V_{ij}$ as well  as
the "breathing mass" $M$ are invariant under a scale transformation
\be
\begin{array}{rcl}
F(z,\vartheta ) & \rightarrow & F_{\beta }(z,\vartheta )\equiv F(\beta
z,\vartheta ),\\
\Theta (z,\vartheta ) & \rightarrow & \Theta _{\beta }(z,\vartheta )\equiv
\Theta (\beta z,\vartheta ).
\label{eq3.7}
\end{array}
\ee
Using the Ansatz (\ref{eq1.2}) with (\ref{eq2.5a}, \ref{eq2.5b})
one obtains the following expressions for nonzero components of
the inertia tensors (\ref{eq3.6a}-\ref{eq3.6c}):
\begin{eqnarray}
\lefteqn{u_{11} ={\pi \over 2f_{\pi }}\int ^{\infty }_{0}dz
z^{2}\int^{\pi }_{0}
d\vartheta
\sin \vartheta (1+\cos ^{2}\Theta )\sin ^{2}F,}
\label{eq3.8a} \\
\lefteqn{u_{33} ={\pi \over f_{\pi }}\int^{\infty }_{0}dz z^{2}\int^{\pi }_{0}
d\vartheta \sin \vartheta  \sin ^{2}\Theta  \sin ^{2}F,}
\label{eq3.8b} \\
\lefteqn{w_{11} =\delta _{1m}{\pi \over 2f_{\pi }}\int^{\infty }_{0}dz z^{2}
\int^{\pi }_{0}d\vartheta \sin \vartheta \left[{\partial \Theta \over
\partial \vartheta }+\cot \vartheta \cos \Theta \sin \Theta \right]
\sin ^{2}F,}
\label{eq3.8c} \\
\lefteqn{v_{11} ={\pi \over 2f_{\pi }}\int^{\infty }_{0}dz z^{2}\int^{\pi }_{0}
d\vartheta \sin \vartheta \left\{\left[\left({d\Theta \over d\vartheta }\right)
^{2}+\right. \right. } \nonumber \\
& & \left. \left. m^{2}\cot ^2 \vartheta \sin ^2 \Theta \right]\sin ^2 F+
{\left({\partial F\over \partial \vartheta }\right)}^2\right\},
\label{eq3.8d} \\
\lefteqn{u_{22} =u_{11},\ w_{22}=w_{11},\ v_{22}=v_{11},}
\label{eq3.8e} \\
\lefteqn{v_{33} =mw_{33}=m^{2}u_{33},}
\label{eq3.8f} \\
\lefteqn{u_{ij} =w_{ij}=v_{ij}=0 \ \ {if}\ i\neq j.}
\label{eq3.8g}
\end{eqnarray}
Due to the relations (\ref{eq3.8f}) the third components of the  canonically-
conjugate momenta
\begin{eqnarray}
K_{j} & \equiv & U_{ij}a_{i}-W_{ij}b_{i},
\label{eq3.9a} \\
L_{j} & \equiv & -W_{ij}a_{i}+V_{ij}b_{i},
\label{eq3.9b}
\end{eqnarray}
are dependent:
\be
mK_{3}=-L_{3},
\label{eq3.10}
\ee
which is a result of the cylindrical symmetry. Now  the  rotational
part of the Hamiltonian becomes:
\begin{eqnarray}
H_{\hbox{rot}}={V_{11}-W_{11}\over 2D}\vec{K}^{2} & + & {U_{11}-W_{11}\over 2D}
\vec{L}^{2}+{W_{11}\over 2D}(\vec{K}+\vec{L})^{2}+ \nonumber \\
& + & {1\over 2}
\left[{1\over U_{33}} - {V_{11}+m^{2}U_{11}-2mW_{11}\over D}\right]K^{2}_{3},
\label{eq3.11}
\end{eqnarray}
where $D\equiv \left| \begin{array}{cc}
                        U_{11} & W_{11} \\
                        W_{11} & V_{11}
                        \end{array}
                        \right|$.
\indent
The canonical angular momenta $\vec{K}$  and $\vec{L}$  are the  so-called
bodyfixed angular momenta in isospace and in usual space, respectively;
the conventional angular momenta $\vec{T}$  and $\vec{J}$  are defined by
the following rotations of this angular momenta $\vec{K}$  and $\vec{L}$,
respectively:
\be
T_{i}=-D_{ij}(A)K_{j},\ \ J_{i}=-D^{T}_{ij}(B)L_{j}.
\label{eq3.12}
\ee
In the quantum theory a scheme of canonical  quantization  for
the chosen rotational coordinates gives rise to the following commutational
relations for momenta $\vec{K} ,\ \vec{L} ,\ \vec{T}$  and $\vec{J}$:
\begin{eqnarray}
\left[K_{i},K_{j}\right] & = & i\epsilon ^{ijk}K_{k},\
\left[L_{i},L_{j}\right]  =  i\epsilon ^{ijk}L_{k},\
\left[K_{i},L_{j}\right]  =  0,
\label{eq3.13a} \\
\left[T_{i},T_{j}\right] & = & i\epsilon ^{ijk}T_{k},\
\left[J_{i},J_{j}\right]  =  i\epsilon ^{ijk}J_{k},\
\left[T_{i},J_{j}\right]  =  0.
\label{eq3.13b}
\end{eqnarray}
with $\vec{T}^{2}=\vec{K}^{2}$ and $\vec{J}^{2}=\vec{L}^{2}$.\\
\indent
Now we can consider a rotational Hamiltonian $(3.11)$ as an  operator
in the Hilbert subspace defined by  the  physical  constraint
generated by $(3.10)$:
\be
\left(mK_{3}+L_{3}\right)|phys>=0.
\label{eq3.14}
\ee
According to Eq.(\ref{eq3.8c}) $W_{11}=0$ for the solitons with $m\ge 2$
and the rotational Hamiltonian is reduced to
\be
H_{rot}={1\over 2U_{11}}\vec{T}^{2}+{1\over 2V_{11}}\vec{J}^{2}+
{1\over 2}\left({1\over U_{33}}-{m^{2}\over V_{11}}-{1\over U_{11}}\right)
K^{2}_{3}.
\label{eq3.15}
\ee
A special case of this expression with $m=2$ was considered in Ref.7.
The case with $m=1$ must be analized separately.\\
\section{Solitons with m=1. Configurations with half-odd Topological charge}
\indent
For $m=1$ the differential equation  (\ref{eq2.18})  is  reduced  to  the
Legendre equation
\be
{d\over dz}\left[(1-z^{2}){dF_{1}\over dz}\right]+\left[
\mu (\mu +1)+2(-1)^{n+1}\right]F_{1}=0.
\label{eq4.1}
\ee
 From the obvious arguments of finiteness of the solution at $-1\le z\le 1$
one obtains that $F_{1}(z)=AP_{k}(z)$, where $A$ is  some  nonzero  constant,
$P_{k}(z)$ stands for the Legendre polynomial of the order of $k$; $\mu $
is imposed by
\be
\mu (\mu +1)+2(-1)^{n+1}=k(k+1).
\label{eq4.2}
\ee
For $even n-value$ the lowest-order solution  is  independent  of
the azimuth angle: $k=0,\ F_{1}=A,\ \mu =1.$ This means that for such solutions
the spherical symmetry is restored and $V_{11}=W_{11}=U_{11}=W_{33}$.  Although
the  determinant $D$  vanishes,  the  ratios ${\ds {(V_{11}-W_{11})}\over
\ds D}$ and ${\ds {(U_{11}-W_{11})}\over \ds D}$ are finite. To compensate the
infinitely  rising ratio ${\ds {W_{11}}\over \ds D}$ one has to impose a new
constraint on the physical states:
\be
\left(\vec{K}+\vec{L}\right)|phys>=0
\label{eq4.3}
\ee
instead of the old one (\ref{eq3.14}). So one  obtains the  standard
classification
$j=t$ of the rotative excitations for  the  solitons  with the hedgehog-like
structure.\\
\indent
Because $\mu $ is real and positive quantity, the  solitons  for  the
$odd n-value$ exist only for $k\ge 2$ and have a nontrivial dependence
on the azimuth angle $\vartheta $. This means that quantum solitons with
half-odd  topological charge could not be spherically-symmetrical,  but  have
the cylindrical symmetry.\\
\indent
For the half-odd topological charge all  measurable  quantities
such as energy, components of inertia tensor etc. are  finite.  The
topological analysis of these solitons was done by K. Fujii and two
of us (N.M.C. and A.P.K.) in Ref.1. The chiral field matrix may  be
regarded as a point on the sphere $S^{int}_{3}$ in the internal  space.  But
the  chiral  field  is  not  defined  uniquely   at   the   origin:
$U(r,\vartheta ,\varphi )\rightarrow i\hat{x}\vec{\tau},\ \hat{x}=
{\ds {\vec{x}}\over \ds r}$, at $r\rightarrow 0$. Calculating measurable
quantities one considers all points $i\hat{x}\vec{\tau}$ of the diameter of
sphere $S^{int}_{3}$ as  one point. This means that the internal space is
reduced to a bundle of spheres $(S_{3}\bigvee S_{3})^{\hbox{int}}$
(see, e.g., Ref.9). On the other hand all  infinite points of the physical
space are reflected to one point $U=1$ of  the internal space, and therefore,
the physical space is  compactificated to $S^{\hbox{phys}}_{3}$. Thus, the
solution for the chiral field is  given by the following mapping:
\be
S^{\hbox{phys}}_{3}\rightarrow (S_{3}\bigvee S_{3})^{\hbox{int}}.
\label{eq4.4}
\ee
When the function $F(y,\vartheta )$ [Eq.(\ref{eq2.5a})]  is  verying  in  the
sector $-{\ds \pi \over \ds 2}\le F\le 0$ the upper sphere of the bundle is
covered.
The lower sphere is  covered  when  the  function $F$  is  verying  in  a
sector $-\pi \le F\le {\ds \pi \over \ds 2}$ and so on. For the half-odd
topological charge the chiral matrix maps the physical space into the
spheres  from  the  bundle $(S_{3}\bigvee S_{3})^{int}$ odd times.\\
\indent
Here we do not consider physical applications of  the  obtained
solitons. We would like only to  mention  that  the  chiral  field,
strictly speaking, could be defined only at the hadron perifery; at
the soliton origin the field matter is determined by color  degrees
of freedom (quarks). In this sense there is  no  problem  with  the
ambiguous behavior of the chiral field at the origin, because  this
region must be "occupied" by the valence quarks and  the  ambiguity
of the chiral field at the origin is an artifact  of  an  effective
theory.\\
\indent
For odd $n$-value $F(y,\vartheta )$ has nontrivial
$\vartheta $-dependence and  in  the expression for the rotative
Hamiltonian the denominator $D$ does not vanish. According to
(\ref{eq3.8c}) $W_{11}\neq 0$ if $m=1$ and, therefore, all  terms
in $(3.11)$ give nonzero contribution to the rotational energy. Moreover,
there appears a new conserved quantity $\vec{\zeta}^{2}=(\vec{K}+\vec{L})^{2}$.
It is obvious that  the  commutation  relations  between  components  of  the
$\zeta $-operator are those for angular momentum:
\be
\left[\zeta _{i},\zeta _{j}\right]=i\epsilon _{\hbox{ijk}}\zeta _{k}.
\label{eq4.5}
\ee
So, for the soliton with half-odd topological charge the  state  is
gi\-ven by
\begin{eqnarray}
|j,t,\zeta ,j_{3},t_{3},k_{3}>, \nonumber
\end{eqnarray}
where $j, t$ and $\zeta $ are
soliton spin,  isospin and "new" momentum, respectively; $j_{3}$ and $t_{3}$
are $z$-components of the spin and isospin; $k_{3}=-t,-t+1,\ldots,t$ is an
eigenvalue  of  the operator $K_{3}$. Due to the constraint (\ref{eq3.14}) the
state  vector  is  an eigenstate of the $\zeta _{3}$-operator with zero
eigenvalue.  Therefore  the quantum number $\zeta $ is integer and is
constrained by
\be
|j-t|\le \zeta \le j+t,
\label{eq4.6}
\ee
and the soliton spin and isospin may be either both integer or both
half-odd.\\
\section{Conclusions and discussion}
\indent
In the above consideration we have investigated  solitons  with
the cylindrical symmetry in the nonlinear $\sigma $-model. The stabilization
of the solitons is provided by quantization  of  the  collective
breathing coordinate. The system of two coupled differential  equations in
partial derivatives for the chiral field is  obtained.  An
analysis of this equations shows that at the origin the chiral angle must be
$\pi /2$ multiplied by any integer number $n$ (see  Eq.(\ref{eq2.15})).
This gives, additionally to the "standard"  solitons  with  integer
topological charge, solitons with half-odd topological charge.  The
latter could not exist in the model, proposed by Skyrme [4],  where
soliton stabilization is guaranteed by {\em ad hoc} added term to the
Lagrangian density of the nonlinear $\sigma $-model (\ref{eq1.1}). It must be
stressed here that the quantum solitons with half-odd topological charge
do not appear for spherically-symmetric Ansatz  of  Refs.1-3.  This
means that for symmetries, which are more general than the  spherical one,
the spectrum of the quantum solitons  is  more  rich  than
that of the static solitons of the model with Skyrme  stabilization term.\\
\indent
Quantization of rotative coordinates for spatial and isospatial
rotations gives appropriate quantum numbers of the soliton. The soliton
with integer topological number is characterized by the standard set of
quantum numbers (spin and isospin). At the  same  time,
the solitons with the ${\ds 1\over \ds 2}$-topological charge are shown
to  be characterized by spin, isospin and by some additional "momentum".  The
operator for this additional momentum commutes with the soliton Hamiltonian,
as well as with its spin and isospin operators.\\
\indent
Between unsolved problems we would like to  mention  here  some
physical applications of the obtained solutions, statistics of  the
solitons with half-odd topological charge  and  numerical  calculations.
We hope to discuss some of them in our further publications.\\
\indent
The authors would like to express their thanks to Professor
K.~Fujii (Hokkaido University) for a number of fruitful discussions.

\end{document}